\documentclass[aps,pre,twocolumn,floatfix,a4]{revtex4-1}
\usepackage{color}
\usepackage{latexsym,amssymb,amsmath,gensymb}
\usepackage{graphicx,dcolumn,bm}
\def\geqsim{\lower.73ex\hbox{$\sim$}\llap{\raise.4ex\hbox{$>$}}$\,$}
\def\leqsim{\lower.73ex\hbox{$\sim$}\llap{\raise.4ex\Shbox{$<$}}$\,$}
\newcommand{\p}{\partial}
\newcommand{\ov}{\overline}
\newcommand{\half}{\frac{1}{2}}
\newcommand{\thalf}{{\textstyle{\frac{1}{2}}}}

\newcommand{\bos}{\boldsymbol}
\newcommand{\tbf}{\textbf}
\newcommand{\tit}{\textit}

\newcommand{\mbf}{\mathbf}

\newcommand{\beq}{\begin{equation}}
\newcommand{\eeq}{\end{equation}}
\newcommand{\bea}{\begin{eqnarray}}
\newcommand{\eea}{\end{eqnarray}}
\newcommand{\barr}{\begin{array}}
\newcommand{\earr}{\end{array}}
\newcommand{\bean}{\begin{eqnarray*}}
\newcommand{\eean}{\end{eqnarray*}}
\newcommand{\bei}{\begin{itemize}}
\newcommand{\eei}{\end{itemize}}
\newcommand{\ben}{\begin{enumeration}}
\newcommand{\een}{\end{enumeration}}
\newcommand{\nn}{\nonumber}

\newcommand{\lam}{\lambda}

\newcommand{\lt}{\left}
\newcommand{\rt}{\right}
\newcommand{\ep}{\epsilon}
\newcommand{\mb}{\mbox}
\definecolor{navyblue}{rgb}{.05,0,.55}

\begin{document}

\title{Models of Membrane Electrostatics}

\author{Kevin Cahill}
\email{cahill@unm.edu}
\affiliation{Biophysics Group,
Department of Physics \& Astronomy,
University of New Mexico, Albuquerque, NM 87131}
\affiliation{Physics Department, Fudan University,
Shanghai, China 200433}

\date{\today}

\begin{abstract}
I derive formulas for the electrostatic potential of a charge in or
near a membrane modeled as one or more dielectric slabs lying between two 
semi-infinite dielectrics.  One can use these formulas in
Monte Carlo codes to compute the distribution of ions near cell
membranes more accurately than by using Poisson-Boltzmann theory or
its linearized version.  Here I use them to discuss the electric field
of a uniformly charged membrane, the image charges of an
ion, the distribution of salt ions near a charged membrane, the
energy of a zwitterion near a lipid slab, and the effect
of including the phosphate head groups as thin layers of high
electric permittivity.  
\end{abstract}

\maketitle

\section{Cell Membranes
\label{Cell Membranes}}

The plasma membrane of an animal cell
and the membranes of the endoplasmic reticulum,
the Golgi apparatus, the endosomes, 
and other membrane-enclosed
organelles are lipid bilayers about 5-nm thick
studded with proteins.
The lipid constituents are mainly phospholipids,
sterols, and glycolipids.
\par
Of the four main phospholipids in membranes,
three---phosphatidylethanolamine (PE),
phosphatidylcholine (PC), and
sphingomyelin (SM)---are neutral,
and one, phosphatidylserine (PS), 
is negatively charged.
In a living cell, PE and PS 
are mostly in the cytosolic layer
of the plasma membrane;
PC and SM are mostly in the 
outer layer~\cite{Zwaal1999, *MBoC4587};
and the
electrostatic potential of the
cytosol is 20 to 120 mV lower
than that of the extracellular environment.
\par
After pioneering work by
Gouy, Chapman, and
Wagner~\cite{Gouy1910, *Chapman1913, *Wagner1924},
and by Onsager and Samaras~\cite{Onsager1934},
many scientists have studied
the electrical properties of
cell membranes~\cite{Parsegian1969, *Bamberg1969, *Neumcke1969, *Neumcke1970, *Brown1974, *Parsegian1975, *Halle1980,
  *McLaughlin1989, *Netz2000, *Flores-Mena2001, *Allen2003,
  *Levin2006, *Kim2007, *Parikh2008, *Frydel2011,
  *Andelman2011, Levin2011}\@.
This paper presents the exact 
electrostatic potential due 
to a charge in or near a
membrane in the continuum limit
in which the membrane is taken to be
one or more dielectric slabs lying between
two different infinite dielectric media.
Because of the superposition principle,
this monopole potential also gives
the multipole potential 
due to any array of charges
in or near a membrane.
\par
One can use these formulas
to simulate the interactions of ions
with other ions and with fixed charges
near membranes while modeling
water and other
neutral molecules as bulk media.
For instance, 
one can use them
in Monte Carlo simulations
to compute
the behavior of salt ions and protons
in water 
near neutral or charged membranes
even in the presence of fixed charges
of arbitrary geometry.
This method is more accurate than
the Poisson-Boltzmann
mean-field approximation and much more
accurate than
its linearized version~\cite{Jordan2010}\@.
These formulas also provide a context for
and a check on all-atom computer 
simulations~\cite{Jordan2010, Wilson1996, *Allen2007}\@.
\par
As early as 1924, Wagner~\cite{Wagner1924}
noted that an ion in water near a lipid
slab induces image charges that repel the ion. 
No mean-field theory can describe this simple
effect.
But work-arounds are available for the 
Poisson-Boltzmann 
theory~\cite{Levin2011}\@. 
\par
Formulas for the electrostatic potential of
a charge in or near a cell membrane 
modeled as a single slab
are derived in 
section~\ref{The Potential of a Charge 
in or near a Lipid Bilayer}\@.
As pedagogical illustrations of their utility,
I use them to compute the 
electric field of a
charged membrane in 
section~\ref{A Surface Charge on a Membrane}
and the response of bound charge to an ion in
section~\ref{Image Charges}\@.
In section~\ref{Ions near a Charged Membrane},
I use them to simulate 
the distribution of salt ions
near a charged membrane.
I discuss the Debye layer in 
section~\ref{Validity of the Debye Layer}
and the energy of a zwitterion 
near a lipid slab
in section~\ref{A Zwitterion}\@. 
In section~\ref{Several Dielectric Layers},
I calculate the potential of a charge near
a membrane modeled as several dielectric layers
of different permittivities between two
different semi-infinite dielectrics.
I use this analysis in 
section~\ref{Three Dielectric Layers}
to model a phospholipid bilayer
as a lipid layer bounded by two layers
of head groups
of high electric permittivity.
The phosphate head groups
cause the membrane to attract
rather than to repel ions.
I summarize the paper in 
section~\ref{Summary}\@.

\section{The Potential of a Charge 
in or near a Lipid Bilayer
\label{The Potential of a Charge 
in or near a Lipid Bilayer}}

In electrostatic problems,
Maxwell's equations reduce to Gauss's law
\(\bos{\nabla \cdot \mbf{D}} = \rho_f\)
which relates the divergence
of the electric displacement \(\mbf{D}\)
to the free-charge density \( \rho_f \)
(not including the polarization of the medium), 
and the static form of Faraday's law
\(\bos{\nabla \times \tbf{E}} = 0\)
which implies that the electric
field \( \mbf{E} \) is the gradient 
of an electrostatic potential
\(\mbf{E} = - \bos{\nabla} V\)\@. 
\par
Across an interface with normal vector \( \mbf{\hat n} \)
between two dielectrics,
the tangential component of the
electric field is continuous
\beq
\mbf{\hat n} \times 
\lt( \mbf{E}_2 - \mbf{E}_1 \rt) = 0
\label {nx(E2-E1)=0}
\eeq
while the normal component
of the electric displacement
jumps by the surface density \( \sigma \)
of free charge
\beq
\mbf{\hat n} \cdot
\lt( \mbf{D}_2 - \mbf{D}_1 \rt) = \sigma.
\label {dD = sigma}
\eeq
In a linear dielectric,
the electric displacement \(\bos{D}\)
is the electric field 
scaled by the permittivity \( \epsilon \) 
of the material
\(
\bos{D} = \epsilon \, \bos{E}
\)\@.
\par
The lipid bilayer
is taken to be flat, 
extending to infinity in
the \(x\)-\(y\) plane,
and of a thickness
\(t \approx 5\)~nm.
The interface between the extracellular
salty water and the lipid bilayer is at \(z = 0\)\@.  
The permittivity \(\ep_\ell\)
of the lipid bilayer is about
twice that of the vacuum
\(\epsilon_\ell \approx 2 \ep_0\);
those of the extracellular environment \( \ep_w \)
and of the cytosol \( \ep_c \)
are about 80 times \(\ep_0\)\@.
\par
The potential of a charge \(q\)
at a point \((0,0,h)\) on the \(z\)-axis
is cylindrically symmetric,  
and so Bessel functions are useful here.
In cylindrical coordinates
with \(\rho = \sqrt{x^2 + y^2}\),
the functions \(J_n(k\rho) \, e^{in\phi} \, e^{\pm k z}\)
form a complete set of solutions of 
Laplace's equation, but due
to the azimuthal symmetry, we  only
need the \(n=0\) functions
 \(J_0(k\rho) \, e^{\pm k z}\)\@.
We will use them and 
the relation~\cite{Schwinger1998.16}
\beq
\frac{1}{\sqrt{\rho^2 + (z - h)^2}}
= \int_0^\infty \! dk \, J_0(k\rho) \, e^{- k |z -h|}
\label{Schwinger}
\eeq
to represent the potential
of point charge at \((0,0,h)\)\@.
\par
If the charge \(q\) is at \((0,0,h)\)
in the water above the membrane (\(h > 0\)),
then we may write the potentials in the 
extracellular water \(V^w_w\),
in the lipid membrane \(V^w_\ell\), and 
in the cytosol \(V^w_c\) as
\bea
V^w_w(\rho,z) & = & 
\int_0^\infty \! dk \, J_0(k\rho) 
\lt[\frac{q}{4\pi \ep_w} e^{- k |z -h|}
+ u(k) \, e^{-kz} \rt] \nn\\
V^w_\ell(\rho,z) & = & 
\int_0^\infty \! dk \, J_0(k\rho) 
\lt[m(k) \, e^{k z}
+ f(k) \, e^{-kz} \rt] \nn\\
V^w_c(\rho,z) & = & 
\int_0^\infty \! dk \, J_0(k\rho) \, d(k) \, e^{k z}.
\label {J exps for q in w}
\eea
Imposing the constraints 
(\ref{nx(E2-E1)=0} \& \ref{dD = sigma}),
writing \(u(k)\) as \(u\), 
\(m(k)\) as \(m\), and so forth, and
setting \(\beta \equiv qe^{-kh}/4\pi \ep_w\)
and \(y = e^{2kt}\), we get the four equations 
\bea
m + f - u & = & \beta \nn\\
\ep_\ell m - \ep_\ell f + \ep_w u 
& = & \ep_w \beta \nn\\
\ep_\ell m - \ep_\ell y f - \ep_c d 
& = & 0 \nn\\
m + y f - d & = & 0.
\label {4eqs for w}
\eea
In terms of the abbreviations 
\bea
\ep_{w\ell} & = & \half \lt( \ep_w + \ep_\ell \rt)
\quad \mbox{and} \quad
\ep_{c\ell} = \half \lt(\ep_c + \ep_\ell\rt) \nn\\
p & = & \frac{\epsilon_w - \epsilon_\ell}
{\epsilon_w + \epsilon_\ell}
\quad \mbox{and} \quad
p' = \frac{\epsilon_c - \epsilon_\ell}
{\epsilon_c + \epsilon_\ell}
\label {p and p' t}
\eea
their solutions are
\bea
u(k) & = &  \beta \, \frac{p - p'/y}{1 - p p'/y} \nn\\
m(k) & = & \beta \, \frac{\ep_w}{\ep_{w\ell}} \,
\frac{1}{1 - p p'/y} \nn\\
f(k) & = & \mb{} - \beta \, \frac{\ep_w}{\ep_{w\ell}} \,
\frac{p'/y}{1 - p p'/y} \nn\\
d(k) & = & \beta \, 
\frac{\ep_w\ep_\ell}{\ep_{w\ell} \ep_{c\ell}} \,
\frac{1}{1 - p p'/y} .
\label {solutions for charge in water}
\eea
\par
Inserting these solutions
into the Bessel expansions (\ref{J exps for q in w})
for the potentials, expanding their denominators
\beq
\frac{1}{1 - p p'/y} = 
\sum_0^\infty (pp')^n \, e^{-2nkt}
\label {expansion of denominator}
\eeq
and using the integral (\ref{Schwinger}),
we find that the potential \(V^w_w(\rho,z)\)
in the extracellular water 
due to a charge \(q\) at \((0,0,h)\) in 
that water is
\bea
V^w_w(\rho,z) & = & 
\frac{q}{4\pi \epsilon_w}
\lt(\frac{1}{r}
+ \frac{p}{\sqrt{\rho^2 + (z+h)^2}} 
\rt. \label {Vw q w}\\
& & \mbox{} -  \lt. 
p'\lt( 1 - p^2\rt)
\sum_{n=1}^\infty
\frac{(p p')^{n-1} }
{\sqrt{\rho^2 + (z + 2nt + h)^2}} \rt)
\nn
\eea
in which \(r =\sqrt{\rho^2 + (z-h)^2}\) 
is the distance to the charge \(q\), and
\(\sqrt{\rho^2 + (z+h)^2}\) 
is the distance to the principal
image charge \(pq\)\@.
Similarly, the potential \(V^w_\ell\)
in the lipid bilayer is
\bea
V^w_\ell(\rho,z) & = &
\frac{q}{4\pi \epsilon_{w\ell}}
\, \sum_{n=0}^\infty
(p p')^n \lt(
\frac{1}{\sqrt{\rho^2 + (z-2nt-h)^2}} \rt. \nn\\
& & \lt. \mbox{} 
- \frac{p'}{\sqrt{\rho^2 + (z+2(n+1)t+h)^2}}\rt)
\label {Vl q w}
\eea
and the potential \(V^w_c\)
in the cytosol is
\beq
V^w_c(\rho,z) = \frac{q \, \ep_\ell}
{4\pi \ep_{w\ell}\ep_{c \ell}}
\! \sum_{n=0}^\infty
\frac{(p p')^n}{\sqrt{\rho^2 + (z-2nt-h)^2}}.
\label {Vc q w}
\eeq
\par
If the charge \(q\) is in the lipid bilayer
at \((0,0,h)\) with \( - t < h < 0 \),
then the Bessel representations of the
potentials are
\bea
V^\ell_w(\rho,z) & = & 
\int_0^\infty \! dk \, J_0(k\rho) 
\, u(k) \, e^{-kz}  \nn\\
V^\ell_\ell(\rho,z) & = & 
\int_0^\infty \! dk \, J_0(k\rho) 
\lt[ \frac{q}{4\pi \ep_\ell} e^{- k |z -h|}
\rt. \nn\\
& & \qquad \qquad \lt.
+ m(k) \, e^{k z} + f(k) \, e^{-kz} 
\frac{}{} \rt] \nn\\
V^\ell_c(\rho,z) & = & 
\int_0^\infty \! dk \, J_0(k\rho) \, d(k) \ e^{k z}.
\label {J exps for q in l}
\eea
With \(\gamma = q e^{kh}/4\pi \ep_\ell\)
and \(x = e^{-2kh}\),
the Maxwell constraints 
(\ref{nx(E2-E1)=0} \& \ref{dD = sigma})
give us the equations
\bea
u - m - f & = & \gamma \nn\\
\ep_w \, u + \ep_\ell m - \ep_\ell \, f 
& = & \ep_\ell \gamma \nn\\
d - m - y \, f & = & 
x \, \gamma \nn\\
\ep_c \, d - \ep_\ell \, m - \ep_\ell \, y \, f 
& = & \ep_\ell x \, \gamma .
\label {4eqs for l}
\eea
whose solutions are
\bea
u(k) & = &  \frac{\ep_\ell \beta} 
{\ep_{w\ell}} \, \frac{1 - p' x/y}{1 - p p'/y} \nn\\
m(k) & = &  \mb{} - p \beta 
\, \frac{1 - p' x/y}{1 - p p'/y} \nn\\
f(k) & = & \mb{} - \frac{p'\beta}{y} \,
\frac{x-p}{1 - p p'/y} \nn\\
d(k) & = & \frac{\ep_\ell \beta}{\ep_{c\ell}} \,
\frac{x-p}{1 - p p'/y} .
\label {solutions for charge in lipid}
\eea
\par
After putting these solutions into
the Bessel expansions (\ref{J exps for q in l})
and using the integral (\ref{Schwinger})
and the denominator 
sum (\ref{expansion of denominator}),
one finds that the potential \(V^\ell_w\) in the
extracellular water due to a charge
\(q\) at \((0,0,h)\) in the lipid bilayer is
\beq\begin{split}
V^\ell_w(\rho,z) & = 
\frac{q}{4\pi \epsilon_{w\ell}}
\!\lt[ \sum_{n=0}^\infty \!
\frac{(p p')^n}
{\sqrt{\rho^2 + (z+2nt-h)^2}} 
\rt. \label {Vw q l}\\
&  \quad \lt. \mbox{} -
\sum_{n=0}^\infty \!
\frac{p'\,(p p')^n}
{\sqrt{\rho^2 + (z+2(n+1)t+h)^2}} \rt] .
\end{split}
\eeq
The potential \(V^\ell_\ell\) in the lipid bilayer is
\bea
V^\ell_\ell(\rho,z) & = &
\frac{q}{4\pi \epsilon_{\ell}}
\lt[ \sum_{n=-\infty}^\infty \!
\frac{(p p')^{|n|}}{\sqrt{\rho^2 + (z-2nt-h)^2}} \rt. 
\label {Vl q l}\\
& & \qquad \lt. \mbox{} -
\sum_{n=0}^\infty
\frac{p(p p')^n}
{\sqrt{\rho^2 + (z-2nt+h)^2}} \rt.
\nn\\
& & \qquad \lt. \mbox{} -
\sum_{n=0}^\infty
\frac{p'\,(p p')^n}
{\sqrt{\rho^2 + (z+2(n+1)t+h)^2}} \rt] 
\nn
\eea
and that \(V^\ell_c\)  in the cytosol is
\beq\begin{split}
V^\ell_c(\rho,z) & =  \frac{q}
{4\pi \ep_{c\ell}} \!
\lt[ \sum_{n=0}^\infty
\frac{(p p')^n}
{\sqrt{\rho^2 + (z-2nt-h)^2}} \rt.
\\
& \qquad \lt. \mb{} -
\frac{p\,(p p')^n}
{\sqrt{\rho^2 + (z-2nt+h)^2}}
\rt].
\label {Vc q l}
\end{split}
\eeq
\par
Finally and somewhat redundantly, 
we turn to the case
of a charge \(q\) in the cytosol 
at \((0,0,h)\) with \( h < - t \)\@.   
Now the Bessel
expansions of the potentials are
\bea
V^c_w(\rho,z) & = & 
\int_0^\infty \! dk \, J_0(k\rho) \,
u(k) \, e^{-kz} 
\label {J exps for q in c}\\
V^c_\ell(\rho,z) & = & 
\int_0^\infty \! dk \, J_0(k\rho) 
\lt[m(k) \, e^{k z}
+ f(k) \, e^{-kz} \rt] \nn\\
V^c_c(\rho,z) & = & 
\int_0^\infty \! dk \, J_0(k\rho) \,
\lt[\frac{q}{4\pi \ep_w} e^{- k |z -h|} 
+ d(k) \, e^{k z} \rt].
\nn
\eea
The continuity conditions 
(\ref{nx(E2-E1)=0} \& \ref{dD = sigma})
give us the four equations
\bea
u - m - f & = & 0 \nn\\
\ep_w u + \ep_\ell m - \ep_\ell f & = & 0 \nn\\
d - m - y f & = & \mb{} - \beta y \nn\\
\ep_c d - \ep_\ell m + \ep_\ell y f & = & \beta c y
\label {4eqs for c}
\eea
whose solutions are
\bea
u(k) & = & \frac{\ep_\ell \, \ep_c}
{\ep_{w\ell} \, \ep_{c \ell}} \,
\frac{\beta}{1 - p p'/y} \nn\\
m(k) & = & \mb{} - \frac{\ep_c}{\ep_{c\ell}} \,
\frac{p \, \beta}{1 - p p' / y}
\nn\\
f(k) & = & \frac{\ep_c}{\ep_{c\ell}} \,
\frac{\beta}{1 - p p' / y}
\nn\\
d(k) & = & \beta \, \frac{p' y - p}{1 - p p'/y}.
\label {solutions for charge in cytosol}
\eea
\par
Thus using 
(\ref{Schwinger} \& \ref{expansion of denominator}) 
in (\ref{J exps for q in c}),
we find that the potential \(V^c_w\) in the
extracellular water due to a charge
\(q\) at \((0,0,h)\) in the cytosol is
\beq
V^c_w(\rho,z) = \frac{q \, \ep_\ell}
{4 \pi \, \ep_{w\ell} \, \ep_{c\ell}}
\sum_{n=0}^\infty \frac{(p p')^n}
{\sqrt{\rho^2 + ( z + 2nt -h)^2}}.
\label {Vw q c}
\eeq
The potential \(V^c_\ell\) in the lipid bilayer is 
\beq\begin{split}
V^c_\ell(\rho,z) & =  \frac{q}
{4 \pi \, \ep_{c\ell}} 
\lt[ \sum_{n=0}^\infty \frac{(p p')^n}
{\sqrt{\rho^2 + ( z - h + 2nt)^2}} \rt.
\\
& \quad \lt. \mb{} - p \sum_{n=0}^\infty \frac{(p p')^n}
{\sqrt{\rho^2 + ( z + h - 2nt)^2}} \rt].
\label {Vl q c}
\end{split}\eeq
The potential \(V^c_c\)  in the cytosol is
\beq\begin{split}
V^c_c(\rho,z) & =  
\frac{q}{4\pi \epsilon_c}
\lt(\frac{1}{r}
+ \frac{p'}{\sqrt{\rho^2 + (z + h + 2t)^2}} 
\rt. \\
&  \mbox{} -  \lt. 
p\lt( 1 - p^{\prime 2}\rt)
\sum_{n=0}^\infty
\frac{(p p')^n }
{\sqrt{\rho^2 + (z - 2nt + h)^2}} \rt)
\label {Vc q c}
\end{split}\eeq
in which \(r\) is the distance to the
charge and \(\sqrt{\rho^2 + (z + h + 2t)^2}\)
is the distance to the principal image
charge \(p'  q\)\@.
\begin{figure}
\centering
\includegraphics[width=3.4in,height=3.3in]{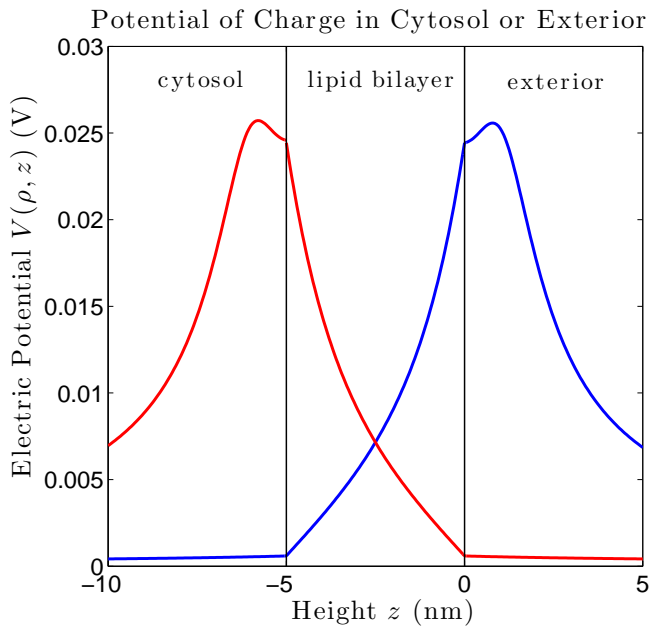}
\caption{(Color online)
The electric potentials \(V^w_w(\rho,z)\),
\(V^w_\ell(\rho,z)\), and \(V^w_c(\rho,z)\)
(\ref{Vw q w}, \ref{Vl q w}, \ref{Vc q w}) and
\(V^c_w(\rho,z)\), 
\(V^c_\ell(\rho,z)\), and \(V^c_c(\rho,z)\)
(\ref{Vw q c}, \ref{Vl q c}, \ref{Vc q c}) 
are plotted (V) for \(\rho = 1\)  
as a function of the height \(z\)
above the phospholipid bilayer 
for a unit charge \(q=|e|\) 
in the cytosol at  
\((\rho,z) = (0,-6)\)   (left curve, blue)
and in the extracellular salty water
at  \((0,1)\)  (right curve, red)\@.
The lipid bilayer is between
\(z = -5\) and \(z = 0\);
distances are in nm; 
and the permittivities are
\(\ep_w = \ep_c = 80 \ep_0\) and
\(\ep_\ell = 2 \ep_0\)\@.}
\label {vh35fig}
\end{figure}

\begin{figure}
\centering
\includegraphics[width=3.4in,height=3.3in]{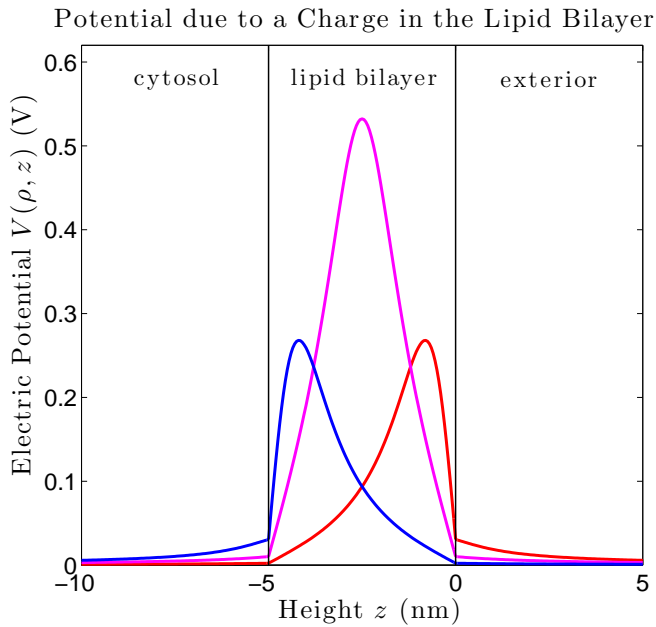}
\caption{(Color online)
The electric potentials \(V^\ell_w(\rho,z)\),
\(V^\ell_\ell(\rho,z)\), and \(V^\ell_c(\rho,z)\)
(\ref{Vw q l}, \ref{Vl q l}, \& \ref{Vc q l})
(V) for \(\rho = 1\)  
as a function of the height \(z\)  
above the phospholipid bilayer 
for a unit charge \(q=|e|\) 
in the phospholipid bilayer at 
\((\rho,z) = (0,\mb{} - 4.5)\)   (left curve, blue), 
\((0,-2.5)\)   (middle curve, magenta), and 
\((0,-0.5)\)   (right curve, red)\@.
Distances and permittivities are
as in Fig.~\ref{vh35fig}\@.}
\label {vh02fig}
\end{figure}

\par
Inasmuch as
\(1-p^2 = \ep_w\ep_\ell/\ep^2_{w\ell}\),
the series (\ref{Vw q w}--\ref{Vc q w})
for the potentials \(V^w_\ell(\rho,z)\),
\(V^w_w(\rho,z)\),
and \(V^w_c(\rho,z)\)
agree with those derived
by the method of image 
charges~\cite{Cahill2010}\@.
The first eight terms of the
infinite series (\ref{Vw q w}--\ref{Vc q w},
\ref{Vw q l}---\ref{Vc q l}, \&
\ref{Vw q c}---\ref{Vc q c})
give the potentials
to within a percent.
They are fast.
\par
The first 1000 terms of the
series (\ref{Vw q w}, \ref{Vl q w}, 
\& \ref{Vc q w}) 
for the potentials \(V^w_\ell(\rho,z)\),
\(V^w_w(\rho,z)\),
and \(V^w_c(\rho,z)\)
(right curve, red)
and (\ref{Vw q c}, \ref{Vl q c}, 
\& \ref{Vc q c}) 
for the potentials \(V^c_\ell(\rho,z)\),
\(V^c_w(\rho,z)\),
and \(V^c_c(\rho,z)\) (left curve, blue)
are plotted in Fig.~\ref{vh35fig} 
(V) for \(\rho = 1\) nm
as a function of the height \(z\) (nm) 
above the phospholipid bilayer 
for a unit charge \(q=|e|\) in the 
extracellular medium at 
\((\rho,z) = (0,1)\) nm (right curve, red)
and in the cytosol at
\((\rho,z) = (0,\mb{} - 6)\) nm 
(left curve, blue)\@.
The errors due to the truncation
of these series 
at 1000 terms are less than 1 part in
\(10^{15}\)\@.
\par
The first 1000 terms of the
series (\ref{Vw q l}, \ref{Vl q l}, 
\& \ref{Vc q l})
for the potentials \(V^\ell_\ell(\rho,z)\),
\(V^\ell_w(\rho,z)\),
and \(V^\ell_c(\rho,z)\)
are plotted in Fig.~\ref{vh02fig} 
(V) for \(\rho = 1\) nm
as a function of the height \(z\) (nm) 
above the phospholipid bilayer 
for a unit charge \(q=|e|\) in the bilayer at 
\((\rho,z) = (0,\mb{} - 4.5)\) (left curve, blue), 
\((0,-2.5)\) (middle curve, magenta), 
and \((0,-0.5)\) nm (right curve, red)\@.
The lipid bilayer extends from \(z= \mb{} - 5\)
to \(z = 0\) nm and is bounded
by thin (black) vertical lines in  
Figs.~\ref{vh35fig} \& \ref{vh02fig}\@.
The cytosol
lies below \(z = \mb{} - 5\) nm.
The relative permittivities 
were taken to be \(\ep_w = \ep_c = 80\)
and \(\ep_\ell = 2\)\@.
\par
Figs.~\ref{vh35fig} \& \ref{vh02fig} 
show that
the potentials fall off sharply
as they cross the lipid bilayer.
The reason for this effect is
a build-up of bound charge 
in the water near the lipid bilayer
due to the high electric permittivities
of the extracellular environment \(\ep_w\)
and of the cytosol \(\ep_c\)\@.
\par
In sections~\ref{Several Dielectric Layers} \& \ref{Three Dielectric Layers}, I will extend this derivation 
to the case of several dielectric slabs.
This generalization will allow us 
to add two layers of high-permittivity
dielectric that will represent the head groups
of the phospholipids.

\section{A Surface Charge on a Membrane
\label{A Surface Charge on a Membrane}}

As a pedagogical application
of the formulas of the
preceding section,
let us consider a uniform 
charge density \(\sigma\)
on the surface of a lipid
bilayer of thickness \(t\)\@.
To avoid minus signs, I will put
the charge density on the 
extracellular leaflet.
After doing the computation,
I will translate the result to 
the case of phosphatidylserine 
on the cytosolic leaflet.
\par
Maxwell's jump equation (\ref{dD = sigma})
tells us that the electric displacement
\(D_w\) in the water differs by
\(\sigma\) from its value \(D_\ell\) 
in the lipid, which in turn is
the same as its value \(D_c\)
in the cytosol.  So we have
two equations
\(D_w = D_\ell + \sigma\)
and \(D_\ell = D_c\)
for three unknowns.
We can use the 
electrostatic potentials of 
section~\ref{The Potential of a Charge 
in or near a Lipid Bilayer}
to resolve this ambiguity. 
\par
The electrostatic potential 
\(V^w_w(\rho,z)\) 
due to a point charge at 
\((0,0,h)\) as given by (\ref{Vw q w})
is equal to the electrostatic potential
\(V^w_w(0,z)\) due to a point
charge at \((\rho,h)\)\@.
Thus setting the height
\(h\) in (\ref{Vw q w}) 
equal to zero, 
and differentiating
with respect to \(z\),
we find for the \(z\)-component
of the electric field at \((0,0,z)\)
due to a charge \(q\)
at \((\rho,0)\)
\beq
E_z(0,z) = \lt. \mb{} -
\frac{\p}{\p z} \, V^w_w(\rho,z)
\rt|_{h=0}
\label{Ew0z}
\eeq
or
\bea
E_z(0,z) & = & 
\frac{q}{4\pi\epsilon_w}
\lt[\frac{(1+p)z}{r^3} 
\rt. \label {Vwdiff}\\
& & \mbox{} -  \lt. 
p'(1-p^2) \sum_{n=1}^\infty
\frac{(p p')^{n-1} \, (z + 2nt)}
{[\rho^2 + (z + 2nt)^2]^{3/2}} \rt]
\nn
\eea
in which \(r = \sqrt{\rho^2 + z^2}\)
and \(z \ge 0\)\@.
Replacing the charge \(q\) by
\(\sigma \, 2 \pi \rho d \rho\)
and integrating over \(\rho\)
from \(\rho = 0\) to \(\rho = \infty\),
we have
\beq
E_z(\sigma) =  
\frac{2 \pi \sigma}{4\pi\epsilon_w}
\lt[ 1
+ p \mbox{} - p'(1-p^2)
\sum_{n=0}^\infty
(pp')^n  \rt].
\label {Esig}
\eeq
The dependence
upon the variables \(z\) and \(t\) 
has dropped out.
Doing the sum
and using the definitions (\ref{p and p' t})
of \(p\) and \(p'\),
we get
\beq\begin{split}
E_z(\sigma) & =  
\frac{\sigma}{2 \epsilon_w}
\lt( 1 + p \mbox{} - p'(1-p^2)
\, \frac{1}{1 - p p'} \rt) \\
& =  
\frac{\sigma}{2 \epsilon_w}
\, \frac{(1+p)(1-p')}{1 - p p'} 
= \frac{\sigma}{\epsilon_w + \ep_c} .
\label {EsigSimple}
\end{split}\eeq
The field points in the \(\bos{\hat z}\)
direction.
\par
In the limiting case
in which the electric permittivity
of the extracellular medium is the
same as that of the cytosol,
\(\ep_w = \ep_c\), the 
electric field is 
\(E_w(\sigma) =  \sigma/2 \ep_w \)\@.
\par
We may apply similar reasoning
to the formula (\ref{Vc q w})
for the potential \(V^w_c(\rho,z)\)
in the cytosol due to a charge
\(q\) in the extracellular
water at a height \(h\) above the membrane.
If we keep in mind that
the quantity \(z-2nt-h\)
is negative, then we find
for the electric field
in the cytosol of 
a surface-charge density
\(\sigma\) at \(h = 0\)
\beq
E_c(\sigma) = \mb{}
- \frac{\sigma}{\ep_w + \ep_c}.
\label{Ecsimple}
\eeq
\par
Since there is no surface charge
between the cytosol and the membrane,
it follows from 
Maxwell's jump 
equation (\ref{dD = sigma}) that 
\(
D_\ell = \ep_\ell \, E_\ell
= D_c = \ep_c \, E_c
\),
and so that the electric field
in the membrane is proportional
to that in the cytosol
\beq
E_\ell(\sigma) = \frac{\ep_c}{\ep_\ell}
E_c(\sigma).
\label {Eell}
\eeq
Our formula (\ref{Ecsimple})
for \(E_c(\sigma)\) now gives
\(E_\ell(\sigma)\) as
\beq
E_\ell(\sigma) = \mb{}
- \frac{\sigma \ep_c}{\ep_\ell (\ep_w + \ep_c)}.
\label{Eellsimple}
\eeq
\par
The jump in the displacement \(D\)
across the layer of surface charge is
\beq
D_w - D_\ell = 
\frac{\sigma \ep_w}{(\epsilon_w + \ep_c)}
+ \frac{\sigma \ep_c}{\ep_w+\ep_c} = \sigma
\label {consistency!}
\eeq
in agreement with
Maxwell's equation (\ref{dD = sigma})\@.
\par
If the layer of surface charge
of density, like that of
phosphatidylserine, lies on
the cytosolic leaflet at \(z = - t\),  
then the electric fields are
\bea
E_w(\sigma) & = & \frac{\sigma}{\ep_w+\ep_c}
\nn\\
E_\ell(\sigma) & = & \frac{\sigma\ep_w}
{\ep_\ell(\ep_w+\ep_c)} 
\label {3 E's for real loc of PSs} \\
E_c(\sigma) & = & \mb{} - \frac{\sigma}
{\ep_w+\ep_c}.
\nn
\eea

\section{Image Charges
\label {Image Charges} }

As noted by Wagner~\cite{Wagner1924},
a charge \(q\) in or near a membrane
polarizes the membrane and 
the surrounding water. 
The potential formulas 
(\ref{Vw q w}--\ref{Vc q w}),
(\ref{Vw q l}--\ref{Vc q l}), and
(\ref{Vw q c}--\ref{Vc q c})
represent these bound charges
as infinitely many mirror charges.
The mirror charges affect the behavior
of ions near an interface
between two dielectrics
in ways that mean-field theories
can't describe.
\par
For instance, a charge \(q\) in the
lipid bilayer induces mirror charges
in the cytosol and in the extracellular
environment.
These induced charges are of opposite
sign, and they attract 
the charge \(q\) in the lipid membrane.
We can be more precise about this
attraction if in the formula (\ref{Vl q l}) 
for \(V^\ell_\ell(\rho,z)\), we use
\(V^{\ell}_\ell(z)\) to represent
the self-potential \(V^\ell_\ell(0,z)\)
without the \(z\)-independent, infinite,
\(n = 0\) term of the first sum
\bea
V^{\ell}_\ell(z) & = &
\frac{q}{4\pi\epsilon_{\ell}}
\lt[ - \frac {\ln ( 1 - p \, p' )}{t}
\rt. \label {mirror effect 1}\\
& & \quad \lt. \mbox{} -
\sum_{n=0}^\infty
\frac{p\,(p p')^n}{|2 \, z - 2n \, t|}
-
\frac{p'\,(p p')^n}
{|2 \, z + 2(n+1) \, t|} \rt] .
\nn
\eea
Keeping only the first term
in each sum and using \(C\) 
for the constant log term, 
we recognize two image charges
\beq
V^{\ell}_\ell(z) \approx
\frac{q}{4\pi \epsilon_{\ell}}
\lt( C - \frac{p}{| 2 \, z |}
- \frac{p'}{| 2 \, z + 2 \, t|} \rt)
\label {mirror effect ex}
\eeq
familiar from freshman physics.
They attract the charge \(q\)
no matter what its sign.
Water is better than lipid
at attracting charges.
\par
Similarly, a charge \(q\) 
in the extracellular water
induces a mirror charge
in the lipid and others in the cytosol.
The mirror charge in the lipid 
is of the same sign and, being
closer, repels the charge \(q\)\@.
We can describe this 
repulsion in terms of the formula (\ref{Vw q w}) 
for \(V^w_w(\rho,z)\) if we use
\(V^w_w(z)\) to mean \(V^w_w(0,z)\)
without the \(z\)-independent, infinite
term \(1/r\)
\beq
V^{w}_w(z)  = 
\frac{q}{4\pi \epsilon_w}
\lt[ \frac{p}{ |2 \, z | }  -  
\frac{\ep_w \ep_\ell}{\ep_{w\ell}^2}
\sum_{n=1}^\infty
\frac{p^{n-1} p^{\prime n}}
{| 2 \, z + 2n \, t |} \rt].
\label {2d mirror charge ex}
\eeq
The first term is the potential
of the textbook mirror charge 
\beq
V^{w}_w(z) \approx
\frac{q}{4\pi \epsilon_w}
\frac{p}{ | 2 \, z | }.
\label {2d mirror charge ex 2}
\eeq
An ion of charge \(q\)
in this potential has an 
energy proportional to \(q^2 \, p\),
which is positive for both
cations and anions.
A lipid membrane therefore repels
both cations and anions; the
water attracts
the ion more than the lipid does.
In a mean-field theory,
such as unpatched Poisson-Boltzmann theory,
every particle responds to the
\tit{same} potential \(V(x)\),
so the force 
\(q \bos{E}(x) = - q \bos{\nabla}V(x)\)
is proportional to the charge \(q\)
of the ion and therefore must
be \tit{opposite} for cations and anions.
Mean-field theories can't describe
why a lipid membrane repels
\tit{both} cations and anions. 

\section{Ions near a Charged Membrane
\label {Ions near a Charged Membrane} }

\begin{figure}
\centering
\includegraphics[width=3.4in,height=3.3in]{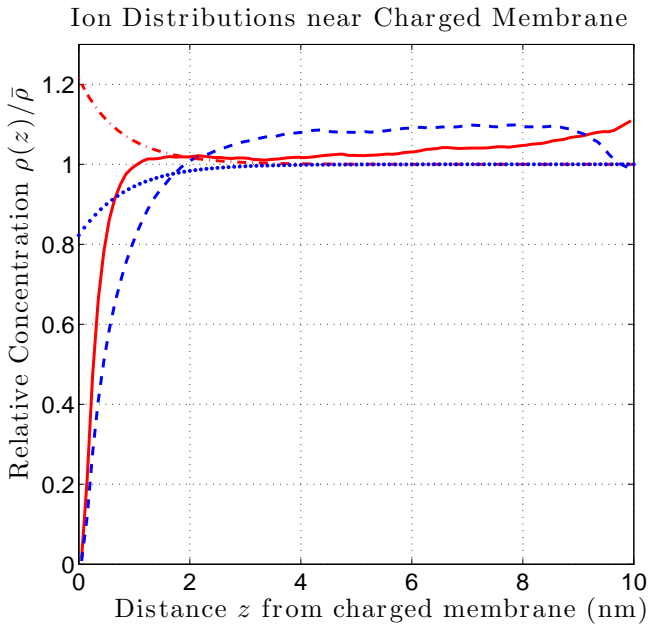}
\caption{(Color online) 
Monte Carlo predictions for the
relative concentrations of potassium 
\(\rho_K(z)/\bar \rho_K\)
(solid, red) and chloride 
\(\rho_{Cl}(z)/\bar \rho_{Cl}\)  
(dashed,blue) ions
at a distance \(z\) (nm) from
the charged cytosolic leaflet 
of a lipid bilayer are plotted along
with the Gouy-Chapman 
predictions (\ref{K Cl G-C}) for the 
normalized potassium \(K_{GC}(z)\) (dot-dash,red) and 
chloride \(Cl_{GC}(z)\) (dots, blue) concentrations.}
\label{gcfig}
\end{figure}

\begin{figure}
\centering
\includegraphics[width=3.4in,height=3.3in]{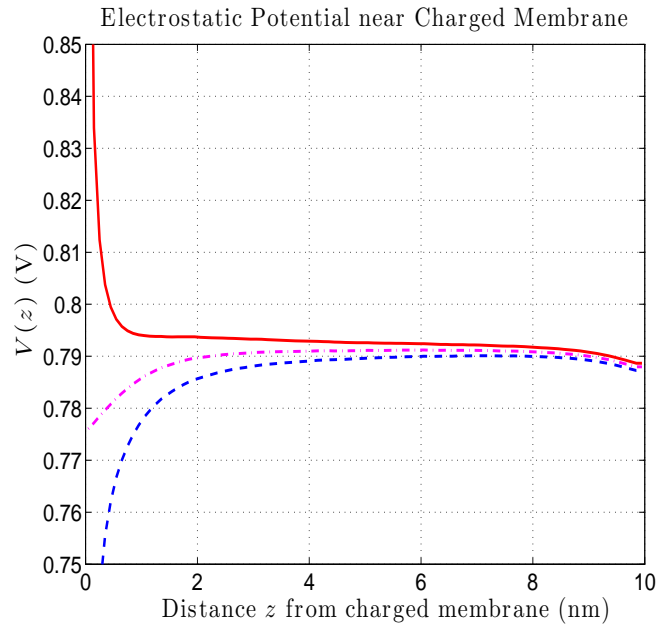}
\caption{(Color online) 
Monte Carlo predictions
for the average total 
electrostatic potential \(V(z)\) 
(V) at a distance \(z\) (nm) from
the charged cytosolic leaflet 
of a lipid bilayer as felt by a 
\(K^+\) (solid, red) and by 
a \(Cl^-\) (dashed, blue) ion.
The potential is that due to the
phosphatidylserines of the cytosolic leaflet, 
the ions of the cytosol,
and the polarization induced by 
the \(K^+\) ion or by the \(Cl^-\) ion.
The dot-dash magenta curve is the 
average potential felt by the ion without the self-potential 
(\ref{2d mirror charge ex}) that represents
the polarization the ion induces.}
\label{potfig}
\end{figure}

\begin{figure}
\centering
\includegraphics[width=3.4in,height=3.3in]{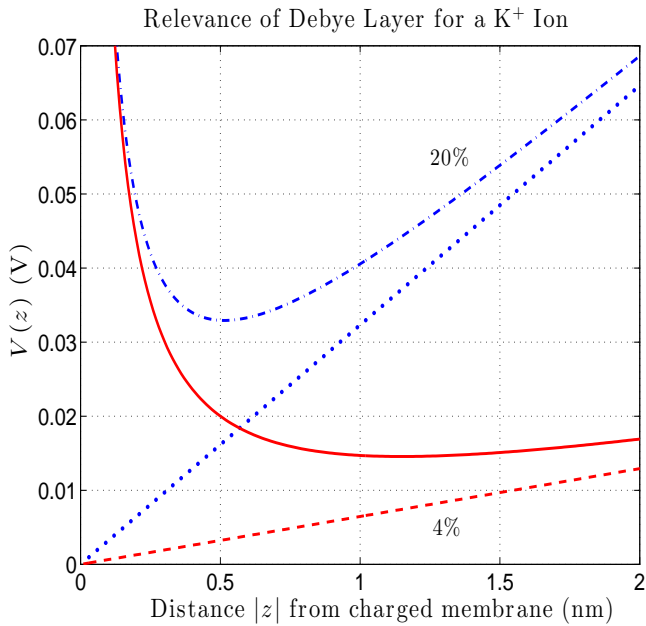}
\caption{(Color online) 
For a potassium ion,
the sum \(V^w_w(z) + V_\sigma(z)\)
of the electrostatic
potential \(V^w_w(z)\) due to 
the induced mirror charges 
(\ref{2d mirror charge ex}) 
and that \(V_\sigma(z) = \mb{}
- \sigma \, |z|/ (\ep_w + \ep_c) \) 
due to the electric field
(\ref{3 E's for real loc of PSs})
of a lipid bilayer 
whose cytosolic leaflet is charged
to a phosphatidylserine mole percent of 
4\% (red, solid curve) or 
20\% (blue, dot-long-dash curve)
is plotted against
the distance \(|z|\) (nm) from the leaflet.
The uncorrected
linear potentials \(V_\sigma(z)\)
of the two surface-charge
densities, 4\% (blue dots)
and 20\% (red dashes),
appear as straight lines.}
\label{debyefig}
\end{figure}

One can use the formulas
of Sec.~\ref{The Potential of a Charge 
in or near a Lipid Bilayer} 
in a Monte Carlo code to compute
the distribution of salt ions
near a charged membrane.
Here I present the result 
of such a simulation of
the distributions of 
potassium and chloride ions
near a membrane of a vesicle
whose inner leaflet contains
phosphatidylserine (PS) at a level
of 4 percent, which is about that
of the plasma membrane of a liver cell.
\par
In the simulation,
I let the potassium and chloride ions
move according to a Metropolis 
algorithm~
within a box whose width
and length were 50 nm and whose
height was 10 nm.
I took the potassium concentration
to be 150 mM so as to allow for 
a 10 mM concentration of sodium ions.
The box contained 2258 \(K^+\) ions.
The bottom of the box was covered by
a uniform negative surface
charge density whose total charge
was \(\mb{} - 143 \, |e|\)
corresponding to 143 phosphatidylserines
at a molar density of 4 \%\@.
I used 2115 \(Cl^-\) ions to make the
whole system neutral; these chloride
ions played the role of the whole
ensemble of anionic
cell constituents. 
\par
To mitigate edge effects,
I surrounded the box with eight
identical boxes into which
I mirrored all 4373 ions.
So there were 39,357 ions
in nine identical boxes.
All the boxes had the same
uniform surface-charge density
due to the presence of the PSs
at a level of 4\,\%\@.
To strictly enforce 
periodic boundary conditions,
one should use 
Ewald sums~\cite{Ewald1921},
but I did not do this
because they would have slowed
the code down and because
liquids are not crystals.
Since the maximum step size 
in the \(z\)-direction was only 1 \AA; 
the error due to using only
eight boxes of mirrored ions
altered the energy difference
\(\Delta E\) of a Monte Carlo move
by less than \( 0.064\, kT\),
usually about  \( 0.016 \, kT\),
far less than the thermal noise.
\par
In my Monte Carlo code~\cite{CahillF902011s},
I used the constant (\ref{EsigSimple}) 
for the electric field \(E_z(\sigma)\) 
of the surface charge of 
phosphatidylserines, the series
(\ref{2d mirror charge ex}) for the
self-potential \(V^w_w(z)\)
arising from the response
of the bound charge to an ion
of charge \(q\),
and the sum (\ref{Vw q w})
for the electrostatic potential
\(V^w_w(\rho,z)\) due to a charge 
in the extracellular water near
a membrane.  
In this way, I took
exact account of the fields
of the charges of the problem
while treating the neutral molecules
of the extracellular environment,
the membrane, and the 
cytosol as bulk dielectric media.
The simulations consisted of eight
separate runs in which 23,000 sweeps
were allowed for thermalization.
Four of the runs collected data
for an additional 50,000 sweeps;
the other four for an additional
9,000 sweeps.
\par
In Poisson-Boltzmann theory~\cite{Nelson2008},
all charges respond to a common
potential \(V\) that obeys
Poisson's equation with a charge
density that respects
the Boltzmann distribution
\beq
\mb{} - \ep \, \triangle V =  \rho_{pf} + \sum_i \rho_{f0i} 
\, e^{\mb{} - q_i V/kT} . 
\label {P-B equation}
\eeq
Here \(V\) is the electrostatic potential,
\(\rho_{pf}\) is a prescribed distribution
of free charge, \(q_i\) is the charge of species \(i\)
and \(\rho_{f0i} \) is its
free-charge density where \(V\) 
vanishes~\cite{Nelson2008}\@.
This non-linear equation
is hard to solve except 
in one-dimensional problems where
the Gouy-Chapman 
solution~\cite{Gouy1910, Chapman1913}
is available~\cite{Nelson2008}\@.
In the present context,
that solution for the potential \(V(z)\)  
is~\cite{Nelson2008}
\beq
V(z) = \mb{} - \frac{2kT}{e} \,
\ln\lt[ \frac{1 + e^{\mb{} - (z + z_0)/\lam}}
{1 - e^{\mb{} - (z + z_0)/\lam}} \rt]
\label {G-C solution}
\eeq
in which \(\lam = 1/\sqrt{8 \pi \ell_B c_\infty}\),
the Bjerrum length is \(\ell_B = e^2/4 \pi \ep_w k T\),
and \(c_\infty\) is the bulk ion concentration
(taken to be the same for potassium
and chloride)\@.
If \(\sigma\) is half the absolute value of
the surface-charge density of the
phosphatidylserines, then
the offset is
\beq
z_0 = \lam \ln \lt[ \frac{e}{2 \pi \ell_B \lam \sigma} \,
\lt( 1 + \sqrt{1 + (2 \pi \ell_B \lam \sigma)^2} \rt) \rt].
\label {offset}
\eeq
The Gouy-Chapman formulas
for the concentrations of the potassium and chloride ions
(normalized to unity where \(V = 0\)) are then
\beq
K_{GC}(z) = e^{\mb{} - eV(z)/kT} \quad
\mb{and} \quad
Cl_{GC}(z) = e^{eV(z)/kT} .
\label {K Cl G-C}
\eeq
\par
In Fig.~\ref{gcfig},
I have plotted my Monte Carlo
predictions for the relative
concentrations of potassium
ions \(\rho_K(z)/\bar \rho_K\)
(red solid curve) and of chloride ions 
\(\rho_{Cl}(z)/\bar \rho_{Cl}\) 
(blue dashed curve)
as functions of the distance \(z\)
from the charged membrane.
The Gouy-Chapman predictions
(\ref{K Cl G-C}) for the
normalized potassium \(K_{GC}(z)\)
(red, dot-dash) and chloride \(Cl_{GC}(z)\)
(blue, dots) concentrations
also are plotted there.
The solid \(K^+\) and dashed \(Cl^-\)
Monte Carlo concentrations correctly drop
sharply for respectively \(z < 1\) 
and \(z < 2\) nm
due to the repulsion by the 
induced image charges
as discussed in 
Section~\ref{Image Charges}\@.
These concentrations are much lower
than the Gouy-Chapman predictions.
The Gouy-Chapman-Poisson-Boltzmann 
potassium concentration actually
rises monotonically when the \(K^+\)
is less than 4 nm from the membrane.
(The behavior of both \(\rho_K(z)/\bar \rho_K\)
and \(\rho_{Cl}(z)/\bar \rho_{Cl}\) for
\(z > 9\) nm is an artifact due to the
absence of ions at \(z > 10\) nm 
in the simulation.)
\par
In Fig.~\ref{potfig}, I have plotted
my Monte Carlo predictions
for the average total potential that
a \(K^+\) ion (solid, red) or 
a \(Cl^-\) ion (dashed, blue) feels 
due to the surface charge of the 
phosphatidylserines, to the other ions,
and to its polarization of the three dielectrics.
The induced bound charge sharply raises
the potential felt by the potassium ion
for \(z < 1\) nm
and lowers that felt by a chloride ion
for \(z < 2\) nm.
The dot-dash magenta curve
represents the potential due 
to the surface-charge layer
of PSs and to
all the ions (including 
the bound charges they induce)
but without the image charges
of equation~(\ref{2d mirror charge ex})\@.
These three potentials differ 
significantly over the whole
range in which the Gouy-Chapman
concentrations differ from their
bulk values.  (The dip in the three
potentials for \(z > 9\) nm
is an artifact due to the
absence of ions at \(z > 10\) nm.)
\par
In the related 
papers~\cite{Cahill2010, Cahill2010iet},
I neglected the self-potential.

\section{Validity of the Debye Layer
\label{Validity of the Debye Layer}}

We have seen in the last two sections
that mean-field theory cannot account
for the behavior of ions near an interface
between two dielectrics with two very different
permittivities.  Does this mean that the
usual interpretation of the Debye layer
is incorrect?
\par
The answer depends upon the 
difference between the two permittivities
and upon the magnitude of the 
surface-charge density.
An image charge is proportional to
the ratio of the difference of
the two permittivities to their sum.
So if the dielectrics have similar
permittivities, then the induced charges
will be weak, and the image-charge correction to 
a  mean-field Debye layer will be small.
Similarly, a high surface-charge density
will dominate
the field due to the induced bound charges 
unless the ions are very close to the
interface.
Thus the validity of the Debye layer
depends on the relative magnitudes
of the self-potential 
\(V^w_w(z) \approx qp/8\pi\ep_w |z|\)
due to the image-charge correction
(\ref{2d mirror charge ex}) and the potential 
\(V_\sigma(z) = - \sigma \, |z|/ (\ep_w + \ep_c) \)
due to the electric field 
(\ref{3 E's for real loc of PSs})
of the phosphatidylserines.
In Fig.~\ref{debyefig}, I have plotted 
for a \(K^+\) ion both
their sum 
\(V^w_w(z) + V_\sigma(z)\)
and the potential \(V_\sigma(z)\)\@.
For the surface-charge density \(\sigma\)
of a 4 mole-percent concentration
of phosphatidylserine, as on the
cytosolic leaflet of a liver cell,
the sum \(V^w_w(z) + V_\sigma(z)\)
(red solid curve)
differs from the 
surface-charge potential \(V_\sigma(z)\)
(red dashed straight line)
for \(z < 2\) nm.
But for a higher mole percent 
of 20\%, the relative difference
between the sum \(V^w_w(z) + V_\sigma(z)\)
(blue dot-long-dash curve) and
\(V_\sigma(z)\) (blue dotted straight line)
is somewhat less.
\par
The importance of the image-charge
correction (\ref{2d mirror charge ex})
rises as the surface-charge density falls.  
It is therefore particularly important
in the case of an uncharged membrane,
such as the outer leaflet of the plasma membrane.

\section{A Zwitterion
\label{A Zwitterion}}

Let us consider a simple model of 
a zwitterionic molecule in salty water
above an uncharged lipid bilayer.
The toy zwitterion is just
a point charge \( q' \)
at \( \bos{r}' \)
and another \( q'' \) 
at \( \bos{r}'' \)\@. 
The charges are separated by
\( \bos{s} = \bos{r}' - \bos{r}'' \)
which makes an angle \( \theta \)
with the vertical \( \bos{\hat z} \)
so that \( \bos{\hat z} \cdot \bos{s} 
= s \cos \theta \) where \(s\)
is the distance between the charges
\( s = \sqrt{ \bos{s} ^2 } \)\@.
The square of the horizontal distance 
between the charges is
\( \rho^2 = (x' - x'' )^2 + (y' - y'')^2 
= s^2 \sin^2 \theta \)\@.
The midpoint of the molecule is
\( \bos{r} = ( \bos{r}' + \bos{r}'' )/2 \)
and its mean height is
\( z = \bos{\hat z} \cdot \bos{r} \)\@. 
The heights of the charges are 
\( z' = z + \thalf s \cos \theta \)
and
\( z'' = z - \thalf s \cos \theta \)\@.
\par
The electrostatic energy \( E'(z') \) 
of the interaction
of the point charge \(q'\) with the
polarization it induces is
\( E'(z') = q'^2 u(z') 
= q'^2 u(z + \thalf s \cos \theta) \) 
in which \( u(z) \) is 
the infinite sum (\ref {2d mirror charge ex})
of image charges
\beq
u(z)  = 
\frac{1}{4\pi \epsilon_w}
\lt[ \frac{p}{ |2 \, z | }  -  
\frac{\ep_w \ep_\ell}{\ep_{w\ell}^2}
\sum_{n=1}^\infty
\frac{p^{n-1} p^{\prime n}}
{| 2 \, z + 2n \, t |} \rt].
\label {2d mirror charge V}
\eeq
Similarly, the function \( u(z) \) gives 
the energy of the interaction
of the other point charge \( q'' \) with the
polarization it induces as
\( E'' ( z'' ) = q''^2 u(z'') 
= q''^2 u(z-\thalf s \cos \theta ) \)\@.
\par
To find the energy 
\( E( \bos{r}',\bos{r}'' ) \) 
of the charge \( q' \)
in the full potential of the charge
\( q'' \) and the energy 
\( E( \bos{r}'',\bos{r}' ) \) 
of the charge \( q'' \)
in the full potential of the charge \( q' \),
we use our formula
(\ref {Vw q w}) for the potential
\( V^w_w(\rho,z) \)\@.
To compute \( E( \bos{r}',\bos{r}'' ) \),
we imagine the charge \( q'' \)
to be at \( (x'', y'', z'') = (0, 0, h) \)
and the charge \( q' \) to be
at \( (\rho, z, 0) \)
in cylindrical coordinates.
Then \( E( \bos{r}',\bos{r}'' ) \)
is \( q' q'' V^w_w(\rho,z)/q \) 
where
\bea
\frac{V^w_w (\rho,z )}
{q}  & = &
\frac{1}{4\pi \epsilon_w}
\lt(\frac{1}{s}
+ \frac{p}{\sqrt{\rho^2 + (z'+z'')^2}} 
\rt. \label {Vw r1 r2 w}\\
& & \mbox{} \!\!\! -  \lt. 
p'\lt( 1 - p^2\rt)
\sum_{n=1}^\infty
\frac{(p p')^{n-1} }
{\sqrt{\rho^2 + (z' + 2nt + z'')^2}} \rt).
\nn
\eea
This interaction is unchanged
when we interchange the locations 
of the two charges, and
\(z' + z'' = 2z\)\@.
So \( E( \bos{r}',\bos{r}'' ) =
E( \bos{r}'',\bos{r}' ) = q'q'' 
v(z, \theta, s) \) where
\bea
v(z, \theta, s ) & = &
\frac{1}{4\pi \epsilon_w}
\lt(\frac{1}{s}
+ \frac{p}
{\sqrt{s^2 \sin^2 \theta + (2 z)^2}} 
\rt. \label {Vw w s theta}\\
& & \!\!\!\!\! \mbox{} -  \lt. 
p'\lt( 1 - p^2\rt)
\sum_{n=1}^\infty
\frac{(p p')^{n-1} }
{\sqrt{ s^2 \sin^2 \theta 
+ ( 2 z + 2nt )^2}} \rt).
\nn
\eea
The total electrostatic energy
of the molecule is then
\beq
\begin{split}
E(z, \theta) & = q'^2 
u(z + \thalf s \cos\theta) +
q''^2 u(z - \thalf s \cos\theta) \\
& \qquad \quad 
\mb{} + 2\, q' \, q'' \, v(z, \theta, s )  .
\label {energy of pol mol}
\end{split}
\eeq
\begin{figure}
\centering
\includegraphics[width=3.4in,height=3.3in]{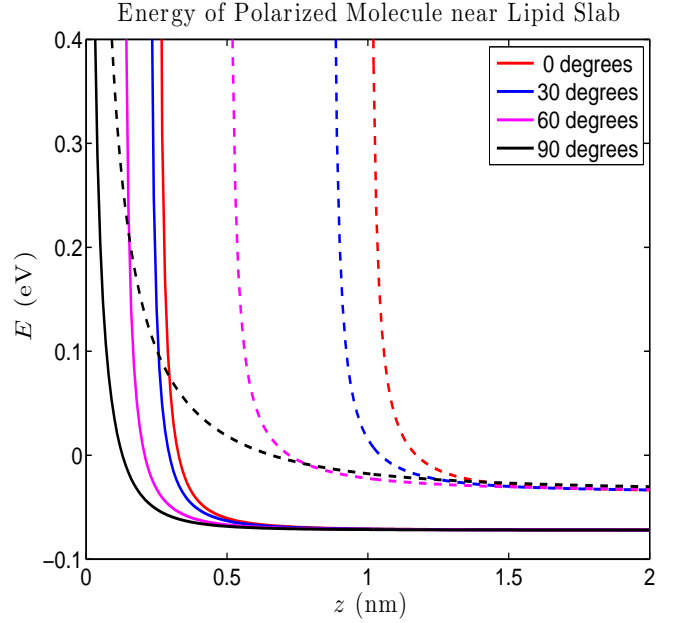}
\caption{(Color online) 
The energy 
(\ref {energy of pol mol} )
of a zwitterion
in salty water a distance \(z\)
from a lipid slab.  The four lower,
solid curves describe a molecule
consisting of a point charge
\( q' =|e| \) separated by 0.5 nm
from a charge \( q'' = - |e| \);
the four upper, dashed curves
are for a molecule
consisting of a point charge
\( q' = 2|e| \) separated by 2 nm
from a charge \( q'' = - |e| \)\@.
Within each quartet the molecules 
from right to left
make angles of 0, 30, 60, 
and 90 degrees with the vertical. }
\label {dipolefig}
\end{figure}
\par
In Fig.~\ref {dipolefig},
I plot the energy 
(\ref {energy of pol mol} ) 
of the model zwitterion when it is
\(z\) nm away from a plasma
membrane considered as a
lipid slab with salty water on both
sides.  The four lower, solid curves
are for a point charge
\( q' =|e| \) separated by 0.5 nm
from a charge \( q'' = - |e| \);
the four upper, dashed curves are for
a point charge
\( q' = 2|e| \) separated by 2 nm
from a charge \( q'' = - |e| \)\@.
The lipid slab repels the molecules.
\par
One may find the energy
of a more realistic zwitterion
by integrating
the two-charge formula
(\ref {energy of pol mol}) 
over a suitable charge distribution.

\section{Several Dielectric Layers
\label{Several Dielectric Layers}}

We may extend our derivation
of the electric potential of
a charge in a material of three
dielectrics to the case of several
dielectrics.  One biological
application is to a phospholipid
bilayer considered as a layer
of lipids bounded by two thin layers
of head groups.  These three layers
with the extracellular water and 
the cytosol pose a five-dielectric problem.
\par
Let us first 
consider the case of 
four dielectric layers
with the charge \(q\)
in the first layer of
permittivity \(\ep_w\)
in the region \(z > 0\)\@.
Instead of the three
potentials (\ref {J exps for q in w}),
we have four
\bea
V^w_w(\rho,z) & = & 
\int_0^\infty \! dk \, J_0(k\rho) 
\lt[\frac{q}{4\pi \ep_w} e^{- k |z -h|}
+ u(k) \, e^{-kz} \rt] \nn\\
V^w_{ 1} (\rho,z) & = & 
\int_0^\infty \! dk \, J_0(k\rho) 
\lt[m_1(k) \, e^{k z}
+ f_1(k) \, e^{-kz} \rt] \nn\\
V^w_{ 2}(\rho,z) & = & 
\int_0^\infty \! dk \, J_0(k\rho) 
\lt[m_2(k) \, e^{k z}
+ f_2(k) \, e^{-kz} \rt] \nn\\ 
V^w_c(\rho,z) & = & 
\int_0^\infty \! dk \, J_0(k\rho) \, d(k) \, e^{k z}
\label {4 exps for q in w}
\eea
in which the first internal layer
of permittivity
\(\ep_{ 1}\) fills the region
\(-t_1 < z < 0\)
while the second internal layer
of permittivity
\(\ep_{ 2}\) fills the region
\(-t_1 -t_2 < z < -t_1\)\@.
The constraints 
(\ref{nx(E2-E1)=0} \& \ref{dD = sigma})
give six equations
\bea
m_1 + f_1 - u & = & \beta 
\nn\\ 
\ep_1 m_1 - \ep_1 f_1 + \ep_w u 
& = & \ep_w \beta 
\nn\\ 
m_1 + y_1 f_1 - m_2 - y_1 f_2  & = & 0 
\nn\\ 
\ep_1 m_1 - \ep_1 y_1 f_1 - \ep_2 m_2 + \ep_2 y_1 f_2  & = & 0 
\nn\\  
m_2 + y_2 f_2 - d  & = & 0 
\nn\\ 
\ep_2 m_2 - \ep_2 y_2 f_2 - \ep_c d 
& = & 0   
\label {6eqs for w}
\eea
in which \(y_1(k)  = \exp(2 k t_1) \) and
\( y_2(k) = \exp(2 k (t_1 + t_2)) \), 
while as in  (\ref{4eqs for w})
the parameter
\( \beta(k) = q \exp(- k h)/4 \pi \ep_w \)
represents the charge at \(z = h > 0\)\@.
The functions \( m_1(k) \) 
and \( f_1(k) \) determine
\( m_2(k) \) and \( f_2(k) \) as
\begin{equation}
\begin{split}
m_2 & = [ ( \ep_2 + \ep_1 ) m_1
+ ( \ep_2 - \ep_1 ) y_1 f_1]/2\ep_2 \\
f_2 & = [( \ep_2 - \ep_1 ) m_1/y_1
+ ( \ep_2 + \ep_1 ) f_1 ]/2\ep_2 .
\label {solution for two layers}
\end{split}
\end{equation}
\par
We now address
the problem of a charge \(q\)
in a semi-infinite
region of permittivity \( \ep_w \)
at a height \(h\) above
\(n\) internal layers of
permittivity \(\ep_i\)
and thickness \( t_i \)
which in turn are above
a semi-infinite
region of permittivity \( \ep_c \)\@.
In the \( i \)th internal layer,
the potential is
\beq
V^w_i (\rho,z) =  
\int_0^\infty \! dk \, J_0(k\rho) 
\lt[m_i(k) \, e^{k z}
+ f_i(k) \, e^{-kz} \rt] 
\label { V_i }
\eeq
while the first and fourth
equations of the set
(\ref {4 exps for q in w})
describe the potentials
in the two semi-infinite regions.
The functions
\( m_i(k) \) and \( f_i(k) \)
determine those
\( m_{i+1}(k) \) and \( f_{i+1}(k) \)
of the next layer
by matrix multiplication.
If 
\beq
p_i = \frac{\ep_{i+1} - \ep_i}
{\ep_{i+1} + \ep_i}  \quad
\mb{and} \quad
\ov \ep_i = \frac {\ep_{i+1} + \ep_i}{2} 
\label {def of p_i}
\eeq
then 
\bea
\begin {pmatrix} m_{i+1} \\
          f_{i+1}
\end {pmatrix}
& =  & \frac{1}{2 \ep_{i+1}}
\begin {pmatrix} \ep_{i+1} + \ep_i &
         (\ep_{i+1} - \ep_i) y_i    \\
          (\ep_{i+1} - \ep_i) / y_i &
           \ep_{i+1} + \ep_i 
\end {pmatrix}
\begin {pmatrix} m_i \nn\\
          f_i
\end {pmatrix} \\
& =  &\frac{\ov \ep_{i}}{ \ep_{i+1} }
\begin {pmatrix} 1 &  p_i y_i    \\
               p_i / y_i & 1
\end {pmatrix}
\begin {pmatrix} m_i \\
          f_i
\end {pmatrix} 
\label {solution}
\eea
in which \(y_0 = 1\) and 
\( y_i = \exp[ 2 k (t_1 + \dots + t_i) ] \)
for \( i > 0 \)\@.
Let us set \( m_0(k) = \beta(k) \)
and \(f_0(k) = u(k)\) as well as
\(m_{n+1}(k) = d(k) \) and
\( f_{n+1}(k) = 0 \)\@.
In these formulas,
\( \ep_{n+1} \) is \( \ep_c \), and
\( \ep_0 \) is \( \ep_w \),
not the permittivity of the vacuum.
If \( E_\ell \) is the product
of \(\ell+1\) of the matrices (\ref {solution})
\beq
E^{(\ell)} = \prod_{i=0}^\ell
\frac{ \ov \ep_i }{ \ep_{i+1} }
\begin {pmatrix} 1 &  p_i y_i    \\
               p_i / y_i & 1
\end {pmatrix}
\label {E}
\eeq
then we have
\beq
\begin {pmatrix} m_{i+1} \\
          f_{i+1}
\end {pmatrix}
= E^{(i)}
\begin {pmatrix} \beta \\
          u
\end {pmatrix}.
\label {i+1 solution}
\eeq
Setting \( i = n \) gives us
\beq\begin {split}
u & = 
- \beta \, E_{2 1 }^{(n)}/E_{2 2}^{(n)} \\
d & = \beta \, \lt( E_{1 1}^{(n)} - 
E_{1 2}^{(n)} E_{2 1 }^{(n)}/E ^{(n)}_{2 2}  
\rt).
\label {u and d}
\end {split} \eeq

\section{Three Dielectric Layers
\label{Three Dielectric Layers}}

A phospholipid bilayer consists
of a layer of phosphate head groups,
a (double) layer of lipids, and a second
layer of phosphate head groups.
In this section, we will apply the formulas 
of section~\ref{Several Dielectric Layers}
to the problem of a charge \(q\)
at a height \(h\) above such a
bilayer.  There are now three slabs
and two semi-infinite regions.
We must compute \( E^{(3)} \)\@.
I get
\beq\begin {split}
E^{(3)}_{2 1} & = 
p_0 + \frac{p_1}{y_1} + \frac{p_2}{y_2}
+ \frac{p_0p_1p_2 y_1}{y_2} \\
& + \frac{p_3}{y_3} 
\lt( 1 + p_0p_1y_1 + p_0p_2y_2
+ \frac{p_1p_2y_2}{y_1} \rt)
\label {E21}
\end {split}\eeq
and
\beq\begin {split}
E^{(3)}_{2 2} & = 
1 + \frac{p_0p_1}{y_1} 
+ \frac{p_0p_2}{y_2}
+ \frac{p_1p_2 y_1}{y_2} \\
& + \frac{p_3}{y_3} \lt(
p_0 + p_1y_1 + p_2y_2
+ \frac{p_0p_1p_2y_2}{y_1}
\rt).
\label {E22}
\end {split}\eeq
\par
\begin{figure}
\centering
\includegraphics[width=3.4in,height=3.3in]{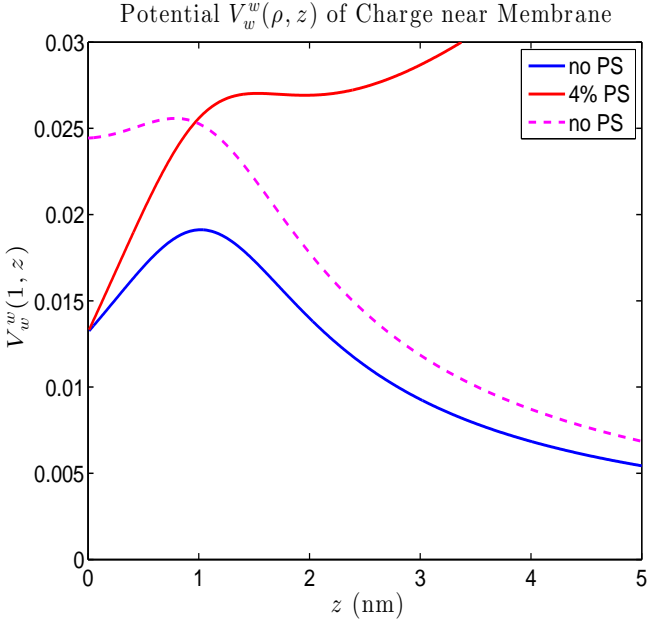}
\caption{(Color online) 
The potential \(V^w_w(\rho,z)\)
(V, \ref{Vww 3slabs}) 
at the point \((\rho,z)\)
due to a charge \(q = |e|\) 
on the \(z\)-axis
at \((0,h)\) above a three-slab
phospholipid bilayer that is
neutral (lowest, solid, blue)
or has a 4 mole-percent layer
of phosphatidylserine on its
cytosolic leaflet (upper, solid, red)\@.
Both \(\rho\) and \(h\)
are 1 nm.
The potential (\ref{Vw q w}) 
of a charge above a 
single neutral lipid
slab without head groups
is plotted as a dashed magenta curve
to illustrate the effect of the layers
of head groups.}
\label {Vww3fig}
\end{figure}
\par
\par
Stern and Feller~\cite{Stern2003},
Nymeyer and 
Zhou~\cite{Nymeyer2008},
and Baker~\cite{Baker2011}
have estimated the relative
electric
permittivity of phospholipid membranes
as being \(1\) from 0 to
10 \AA\ from the center,
4 from 10 to 15  \AA,
180 from 15 to 20 \AA,
210 from 20 to 25 \AA,
and like bulk water beyond
25 \AA\@.
I will approximate their results
by using 2 from 0 to 15 \AA\
and 195 from 15 to 25 \AA\
and will take 
the relative permittivities
of the cytosol and of
the extra-cellular environment
to be 80\@.
These approximations greatly
simplify our formulas
(\ref {E21} \& \ref {E22})
and imply that
\(p_0 = 0.418 = - p_3\) and
\(p_1 = - 0.98 = - p_2\)\@.
The thicknesses of the layers
are \(t_1 = t \equiv 1\) nm,
\(t_2 = 3t\), and \(t_3 = t\),
and so \(y_1 = e^{2kt}\),
\(y_2 = e^{8kt}\), and 
\(y_3 = e^{10kt}\)\@.
With these simplifications,
our formulas (\ref {E21} \& \ref {E22})
reduce to
\beq\begin {split}
E^{(3)}_{2 1} & = 
p_0 + p_1 (e^{-2kt} - e^{-8kt})
- p_0 p_1^2 e^{-6kt}  \\
&  - p_0 e^{-10kt} 
\lt( 1 + p_0p_1 (e^{2kt} - e^{8kt} )
- p_1^2 e^{6kt} \rt) \\
& =  p_0 + p_0 p_1^2 e^{-4kt}  
+ p_1 (1 + p_0^2) (e^{-2kt} - e^{-8kt})
\label {E21sim}\\
&  - p_0 e^{-10kt} 
- p_0 p_1^2 e^{-6kt}
\end {split}\eeq
and
\beq\begin {split}
E^{(3)}_{2 2} & = 
1 + p_0p_1 (e^{-2kt} - e^{-8kt})
- p_1^2  e^{-6kt} \\
& - p_0 e^{-10kt} \lt(
p_0 + p_1 (e^{2kt}  -  e^{8kt} )
- p_0p_1^2 e^{6kt} \rt) \\
& = 1 + p_0^2 p_1^2 e^{-4kt} 
+ 2 p_0p_1 (e^{-2kt} - e^{-8kt}) \\
& - p_0^2 e^{-10kt} 
- p_1^2  e^{-6kt} .
\label {E22sim}
\end {split}\eeq
\par
The key function \(u(k)\)
then is by (\ref {u and d})
the ratio
\bea
u(k) & = & \mb{} 
- \frac{q}{4\pi\ep_w} e^{-kh}
\bigg[ p_0 + 
p_1 (1 + p_0^2)(e^{-2kt} - e^{-8kt})
\nn\\
&  &  \quad \mb{} + p_0 p_1^2 e^{-4kt}  
- p_0 p_1^2 e^{-6kt}
- p_0 e^{-10kt} \bigg] \nn\\
& & \qquad \bigg/
\bigg[ 1 + 2 p_0p_1 (e^{-2kt} - e^{-8kt})
+ p_0^2 p_1^2 e^{-4kt} \nn\\
& & \qquad\qquad \mb{} - p_1^2  e^{-6kt}  
- p_0^2 e^{-10kt} 
\bigg] . \label {u(k)} 
\eea
\par
\begin{figure}
\centering
\includegraphics[width=3.4in,height=3.3in]{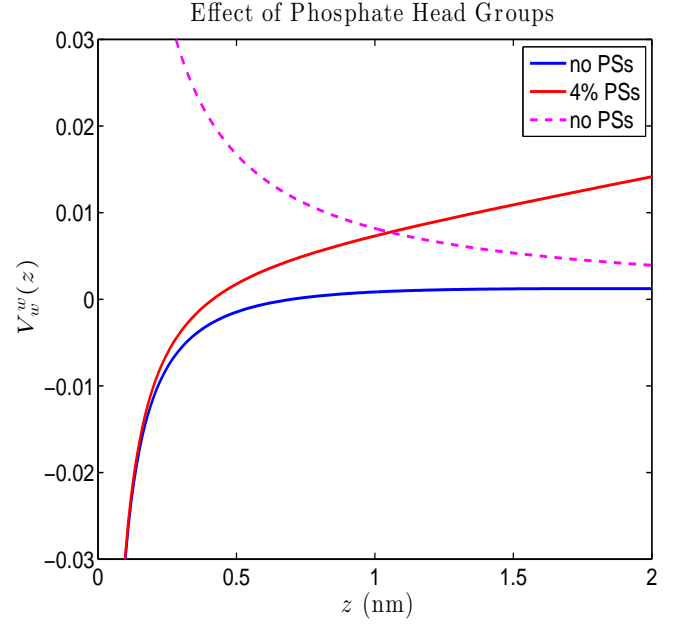}
\caption{(Color online) 
The potential \(V_w^w(z)\)
(V, \ref{Vww(z)}) 
felt by a unit positive charge
at a height \(z\) nm above
a phospholipid membrane
modeled as a lipid slab bounded
by two thin polar slabs.
The lowest (solid, blue) curve 
is for a neutral membrane;
the upper (solid, red) curve is for
a membrane whose cytosolic
leaflet is negatively charged
by phosphatidylserine at
4 mole percent.
Without the outer polar slabs,
the potential (\ref {2d mirror charge ex}) 
is purely repulsive
(dashed magenta)\@.}
\label {Voltfig}
\end{figure}
We can use it and our formula
(\ref {4 exps for q in w})
to write
the potential \(V^w_w(\rho,z)\)
at the point \((\rho,z)\)
due to a charge \(q\) 
on the \(z\)-axis
at \((0,h)\) above a three-slab
phospholipid bilayer as
\beq V^w_w(\rho,z)  =  
\int_0^\infty \! dk \, J_0(k\rho) 
\lt[\frac{q}{4\pi \ep_w} e^{- k |z -h|}
+ u(k) \, e^{-kz} \rt].
\label {Vww 3slabs}
\eeq
\par
Figure~\ref {Vww3fig} plots
this potential \(V^w_w(\rho,z)\)
for \(\rho = 1\) nm
and \(0 \le z \le 5\) nm
for the case of a unit positive
charge \(q = |e|\) 
on the \(z\)-axis at height
\(h = 1\) nm
above a three-slab
phospholipid bilayer that is
neutral (lowest curve, solid, blue)
or has a 4 mole-percent layer
of phosphatidylserine on its
cytosolic leaflet (upper curve, solid, red)\@.
The electric field of the PSs
is taken from 
(\ref{3 E's for real loc of PSs}) to be 
\( E_w(\sigma) = 
\sigma/(\ep_w+\ep_c)\)
in which \(\sigma\) is negative.  
To illustrate the effect of the layers
of head groups with very high
electric permittivity,
I have replotted the potential
(\ref{Vw q w}) of Fig.~\ref {vh35fig} 
due to the same charge but
above a single naked, neutral lipid slab
as a dashed magenta curve.
The head-group dipoles 
lower the potential \( V_w^w(\rho,z)\)
in their vicinity.
\par
We also can use the ratio \( u(k) \)
to compute 
the self-interaction
of a charge \(q\) 
with our three slab
model of the phospholipid
bilayer.  The potential felt
by the charge in the salty water
due to all 
the image charges its presence
induces in the three slabs and
in the cytosol is
\beq
V_w^w(z) = 
\int_0^\infty \!\! 
u(k) \, e^{-kz} \, dk
\label {V 3slabs}
\eeq
in which \(u(k)\) is
the ratio (\ref{u(k)})
but with the height \(h\) of
the charge \(q\) replaced
by \(z\)\@.
The potential felt by the
charge \(q\) at \(z\) then is
\bea
V_w^w(z) & = & - \frac{q}{4\pi\ep_w t} 
\int_0^\infty \!\!\! 
e^{-2kz/t} 
\bigg[ p_0 + p_0 p_1^2 e^{-4k}
- p_0 e^{-10k} 
\nn\\
&  &  \quad \mb{} + 
p_1 (1 + p_0^2)(e^{-2k} - e^{-8k}) 
- p_0 p_1^2 e^{-6k} \bigg] \nn\\
& & \quad \bigg/
\bigg[ 1 +  p_0^2 p_1^2 e^{-4k}
- p_0^2 e^{-10k} 
\label {Vww(z)}\\
& & \qquad \mb{} 
+ 2 p_0p_1 (e^{-2k} - e^{-8k})
- p_1^2  e^{-6k}  
\bigg]  \, dk \nn
\eea
in which the parameter
\(t\) is one nm.
\par
Figure~\ref {Voltfig} plots
the electric potential felt by
an ion of charge \(|e|\) 
near a phospholipid bilayer
modeled as a lipid slab bounded 
by two thin polar slabs.  
The lowest (solid, blue) curve 
is for a neutral membrane;
the upper (solid, red) curve is for
a membrane whose cytosolic
leaflet is negatively charged
by phosphatidylserine at
4 mole percent represented
as in Fig.~\ref{Vww3fig}\@.
The two thin
layers of phosphate head groups
make the potential felt by an ion
near a neutral phospholipid bilayer
attractive rather than repulsive.
Without the outer polar slabs,
the potential (\ref {2d mirror charge ex}) 
is purely repulsive
(dashed magenta)\@.
The head groups attract 
to the membrane ions
that a naked lipid slab
would repel.  If the membrane
has phosphatidylserines on
its cytosolic leaflet,
then it strongly attracts positive ions
that it otherwise would repel.
Apparently the phosphate head groups
facilitate many physiological processes,
such as  the docking of ligands and
the translocation
and endocytosis of positive ions and
cell-penetrating peptides.

\section{Summary
\label{Summary}}
I derived the electrostatic potential 
of a charge in or near a lipid bilayer
in section~\ref{The Potential of a Charge 
in or near a Lipid Bilayer}
and used it in 
section~\ref{A Surface Charge on a Membrane}
to compute the electric field
of a uniformly charged membrane
and
in section~\ref{Image Charges} 
to describe the effects of image charges.
In section~\ref{Ions near a Charged Membrane},
I used the results of 
sections~\ref{The Potential of a Charge 
in or near a Lipid Bilayer}--\ref{Image Charges}
in a Monte Carlo computation of the
distribution of ions near
a charged membrane.
I discussed the validity of the Debye layer in 
section~\ref{Validity of the Debye Layer}
and computed the energy of 
a zwitterion near a lipid slab
in section~\ref{A Zwitterion}\@. 
In section~\ref{Several Dielectric Layers},
I calculated the potential of a charge near
a membrane modeled as several dielectric layers
of different permittivities between two
different semi-infinite dielectrics.
I used this analysis in 
section~\ref{Three Dielectric Layers}
to model a phospholipid bilayer
as a lipid layer bounded by two layers
of head groups
of high electric permittivity.
The phosphate head groups
cause a neutral membrane to attract
rather than to repel ions.

\begin{acknowledgments}
I am grateful to
Nathan Baker, Leonid Chernomordik,
Charles Cherqui,
David Dunlap, Scott Feller, Kamran Melikov,
Michael Wilson, Adrian Parsegian,
Sudhakar Prasad, Harry Stern, and
David Waxman
for helpful conversations; 
to Susan Atlas,
Bernard Becker, Vaibhav Madhok,
Samantha Schwartz, James Thomas, 
and Toby Tolley
for useful comments;
and to the two referees
for constructive criticism.
\end{acknowledgments}
\bibliography{bio,physics}
\end{document}